\documentclass[final,5p,times,twocolumn]{elsarticle}
\journal{Physics Letters B}
\usepackage[T1]{fontenc}
\usepackage{cmap}
\usepackage{amsmath,amssymb,graphicx,booktabs,bm}
\usepackage{placeins}
\usepackage{float}
\usepackage[colorlinks=true,linkcolor=blue,citecolor=blue,urlcolor=blue]{hyperref}
\biboptions{numbers,sort&compress}
\newcommand{\SM}{\mathrm{SM}}

\newcommand{\BR}{\mathcal B}
\newcommand{\ADG}{\mathcal A_{\Delta\Gamma}}
\newcommand{\Smm}{\mathcal S_{\mu\mu}}
\newcommand{\PL}{P_L}
\newcommand{\dd}{\mathrm d}
\makeatletter\def\input@path{{output/}}\makeatother
\graphicspath{{output/}}

\begin{document}
	\raggedbottom
	\begin{frontmatter}
		\title{Resolving complex flavor-amplitude degeneracies in rare $B$ decays: $Z'$ and $S_3$ mediators}
		\author[unisa,infnSal]{Antonio Capolupo}
		\ead{acapolupo@unisa.it}
		\author[unisa,infnSal]{Gabriele Pisacane}
		\ead{gpisacane@unisa.it}
		\author[ssm,infnNa]{Raoul Serao}
		\ead{r.serao@ssmeridionale.it}
		\affiliation[unisa]{organization={Dipartimento di Fisica ``E. R. Caianiello'', Universit\`a degli Studi di Salerno},
			addressline={Via Giovanni Paolo II 132}, city={Fisciano}, postcode={I-84084}, country={Italy}}
		\affiliation[infnSal]{organization={Istituto Nazionale di Fisica Nucleare, Gruppo Collegato di Salerno},
			city={Fisciano}, postcode={I-84084}, country={Italy}}
		\affiliation[ssm]{organization={Scuola Superiore Meridionale},
			addressline={Largo San Marcellino 10}, city={Napoli}, postcode={80138}, country={Italy}}
		\affiliation[infnNa]{organization={Istituto Nazionale di Fisica Nucleare, Sezione di Napoli},
			addressline={Complesso Universitario di Monte S. Angelo, Via Cintia Edificio 6},
			city={Napoli}, postcode={80126}, country={Italy}}
		\begin{abstract}
			We investigate the implications of recent LHCb measurements of rare
			dimuon decays and CP-sensitive $B^0\to K^{*0}\mu^+\mu^-$ angular
			observables for new physics mediated by a heavy $Z'$ boson or a
			scalar triplet leptoquark $S_3$. Combining these measurements with
			neutral-meson mixing, semileptonic and kaon constraints, we examine
			the complementary roles of CP-violating and CP-averaged observables
			in resolving degeneracies among complex flavor amplitudes.
			Particular attention is devoted to the interplay between the
			different semileptonic operator structures and the interference
			terms governing the angular distribution. We then investigate the
			implications of the resulting coupling constraints for
			time-dependent CP violation in rare leptonic $B$ decays and assess
			the sensitivity of these observables to the underlying mediator
			structure.
			The CP-averaged angular data exclude the $S_3$ solution with a nearly
			sign-reversed leptonic decay amplitude, and the calibrated phase profiles
			restrict the tagged CP-asymmetry coefficient $\Smm^s$ in
			$B_s^0\to\mu^+\mu^-$ to $[-0.170,\,0.215]$ for $Z'$ and
			$[-0.145,\,0.351]$ for $S_3$.
		\end{abstract}
		\begin{keyword}
			Rare $B$ decays \sep CP violation \sep Angular observables \sep Flavor physics \sep Leptoquarks \sep $Z'$ bosons
		\end{keyword}
	\end{frontmatter}
	
	\section{Introduction}
	\label{sec:introduction}
	
	Rare beauty-meson decays provide sensitive probes of flavor-changing interactions beyond the Standard Model (SM)~\cite{weak1996,flavor2010}. The evidence for $B^0\to\mu^+\mu^-$ and the improved measurements of $B_s^0\to\mu^+\mu^-$ reported by LHCb in November 2026 sharpen the experimental constraints on these transitions~\cite{lhcb2026}. Their short-distance branching fractions constrain the moduli of the decay amplitudes, so distinct complex new-physics contributions may yield similar rates~\cite{pursuit2017,cpleptonic2018}. Neutral-meson mixing supplies independent phase information through its quadratic dependence on the flavor-changing couplings~\cite{buras2013,time2013,zprime2013}. The extent to which mixing lifts the decay-amplitude degeneracies depends on the mediator and the order at which it contributes to mixing. Capolupo~\cite{capolupo2011cpt} proposed using noncyclic phases in the time evolution of neutral mesons, including $B_s^0$, to probe CPT violation.
	
	The angular distribution of $B^0\to K^{*0}\mu^+\mu^-$ is sensitive to interference among several polarization amplitudes~\cite{transverse2005,egede2010}. CP-averaged observables and CP asymmetries probe complementary combinations of these amplitudes~\cite{angular2009,angularcp2008,optimized2013}. The comprehensive LHCb angular measurements published in 2026, together with their experimental correlations~\cite{angular2026,angular2026data}, and the recent experimental determination of complex Wilson coefficients~\cite{lhcbamplitude2026}, provide new information on the weak phases of the semileptonic transitions. Their interpretation requires correlated form-factor uncertainties and nonlocal hadronic contributions~\cite{charm2010,dispersion2018}.
	
	Large CP-violating phases in the $S_3$ leptoquark scenario and the sensitivity of $b\to s\mu^+\mu^-$ observables to complex couplings have been investigated in Refs.~\cite{s3cp2021,rumble2023,fingerprint2023,cpvectors2026}. Here we follow the evolution of the individual complex-amplitude solutions as the CP-asymmetric and CP-averaged measurements are successively introduced. This allows us to identify which angular interference terms lift the degeneracies of the leptonic and mixing amplitudes and to determine how the resulting coupling regions translate into time-dependent CP observables.
	
	We compare a heavy $Z'$ with left-handed flavor-changing quark couplings and an axial muon interaction to a scalar triplet leptoquark $S_3$~\cite{zprimereview2009,lqreview2016}. We retain complex couplings in both the $b\to d$ and $b\to s$ sectors and the correlated kaon transitions of the scalar flavor structure~\cite{lqkaons2018}. Starting from the updated dimuon measurements, neutral-meson mixing and complementary semileptonic and kaon inputs, we add first the angular CP asymmetries and then the selected CP-averaged observables, retaining the experimental and theoretical correlations. The individual coupling and phase profiles are calibrated with pseudoexperiments~\cite{chernoff1954,selfliang1987}. For $S_3$, the best-fit tagged CP-asymmetry coefficient in $B_s^0\to\mu^+\mu^-$ drops from $0.95$ to $0.145$ with the angular CP asymmetries and to $0.103$ with the full angular data.
	
	The paper is organized as follows. Sections~\ref{sec:models} and~\ref{sec:time} introduce the mediator interactions and the relevant decay, mixing, and temporal observables. Section~\ref{sec:data} describes the experimental inputs, correlated theoretical uncertainties, and statistical inference. Section~\ref{sec:results} presents the coupling constraints and their time-dependent implications, together with the statistical validation and scope of the analysis. Numerical details are collected in the Appendices. Finally, Sec.~\ref{sec:conc} summarizes the main results and discusses their implications for the phenomenology of rare $B$ decays.
	
	\section{Flavor interactions and effective matching}
	\label{sec:models}
	
	In both mediator scenarios, the flavor-changing interactions
	that contribute to rare dimuon decays also modify
	neutral-meson mixing. We consider transitions $b\to q$,
	with $q=d,s$, mediated by a heavy $Z'$ boson or a scalar
	triplet leptoquark $S_3$. To describe their effects on the
	leptonic decay amplitude, we introduce the complex
	parameter $r_q$, defined as the ratio of the new-physics
	contribution to the corresponding Standard Model (SM)
	axial coefficient. Factoring out the
	Cabibbo--Kobayashi--Maskawa (CKM) combination
	$\lambda_t^q=V_{tb}V_{tq}^*$, the effective Hamiltonian
	can be written as
	\begin{equation}
		\begin{aligned}
			\mathcal H_{\rm sl} \supset a_0\lambda_t^q(1+r_q) 
			   (\bar q\gamma_\mu\PL b)
			(\bar\mu\gamma^\mu\gamma_5\mu)+\mathrm{h.c.},
		\end{aligned}
		\label{eq:ham}
	\end{equation}
	where $\PL=(1-\gamma_5)/2$ projects onto left-handed
	quark fields. In this convention, $r_q=0$ corresponds
	to the SM, while its complex phase describes the
	relative weak phase of the additional axial amplitude.
	The SM normalization is $a_0=8.36\times10^{-8}\,n_A\ \mathrm{GeV}^{-2}$,
	with $n_A=1\pm0.015$.
	This normalization follows the nonradiative SM
	prediction of Ref.~\cite{qed2019}.
	
	The same flavor-changing interactions contribute to
	$B_q^0$--$\bar B_q^0$ mixing through the four-quark
	operator
	$Q_{LL}^q=(\bar q\gamma_\mu\PL b)^2$.
	Its hadronic matrix element is parameterized as
	$\langle B_q|Q_{LL}^q(\mu)|\bar B_q\rangle
	=\frac{2}{3}m_{B_q}^2f_{B_q}^2B_q(\mu)$,
	where $f_{B_q}$ is the meson decay constant and
	$B_q(\mu)$ is the scale-dependent bag parameter.
	In the mixing amplitudes, we use the corresponding
	renormalization-group-invariant quantity $\widehat B_q$,
	together with the SM short-distance coefficient
	$\widehat c_\SM=\frac{G_F^2M_W^2}{4\pi^2}\eta_B S_0(x_t)n_M$,
	with $n_M=1\pm0.015$,
	where $G_F$ is the Fermi constant, $M_W$ is the
	$W$-boson mass, $\eta_B=0.551$ accounts for
	short-distance QCD corrections, and $S_0(x_t)$
	is the Inami--Lim loop function, evaluated at
	$x_t=m_t^2(m_t)/M_W^2$ with
	$m_t(m_t)=163.0$ GeV. The factors $n_A$ and
	$n_M$ account for residual short-distance normalization
	uncertainties, each with a relative standard deviation
	of $1.5\%$.
	
	Neglecting new absorptive mixing contributions and
	scalar or pseudoscalar dimuon operators, we introduce
	the decay-amplitude ratio $P_q=1+r_q$ and the
	mixing-amplitude ratio $D_q=M_{12}^q/M_{12}^{q,\SM}=1+\kappa_Xr_q^2$,
	where $M_{12}^q$ is the dispersive mixing amplitude
	and $\kappa_X$ is a real coefficient determined by
	the mediator $X=Z',S_3$ and its short-distance
	matching. Since the flavor-changing couplings may
	be complex, the mixing contribution depends on
	$r_q^2$, rather than $|r_q|^2$.
	
	For either mediator, we denote by $c_A^{q,X}$ and
	$c_{LL}^{q,X}$ the coefficients multiplying,
	respectively, the axial semileptonic operator
	$(\bar q\gamma_\mu\PL b)
	(\bar\mu\gamma^\mu\gamma_5\mu)$
	and the four-quark operator $Q_{LL}^q$ in the
	corresponding new-physics effective Hamiltonian.
	The CKM factors are not included in these
	coefficients.
	
	For the $Z'$ scenario, we consider left-handed
	flavor-changing quark interactions and an axial
	muon coupling,
	\begin{equation}
		\mathcal L_{Z'}=Z'_\mu\left[
		\sum_{q=d,s}g_{qb}\bar q\gamma^\mu\PL b
		+\mathrm{h.c.}
		+g_A\bar\mu\gamma^\mu\gamma_5\mu
		\right],
		\label{eq:Zlag}
	\end{equation}
	where $g_A=(g_R^\mu-g_L^\mu)/2$ is real.
	The muon vector coupling, right-handed
	flavor-changing quark couplings, and independent
	$s\to d$ vector coupling are set to zero.
	
	At energies well below the mediator mass, the
	exchange of a virtual $Z'$ can be described by
	local four-fermion interactions. Integrating out
	the heavy vector field at tree level generates
	an axial semileptonic operator proportional to
	$g_{qb}g_A$ and a left-handed four-quark operator
	proportional to $g_{qb}^2$. The corresponding effective coefficients are
	$c_A^{q,Z'}=g_{qb}g_A/M_{Z'}^2$ and
	$c_{LL}^{q,Z'}=g_{qb}^2/(2M_{Z'}^2)$.
	Identifying $c_A^{q,Z'}=a_0\lambda_t^q r_q$ in
	Eq.~\eqref{eq:ham} then gives
	$g_{qb}=(M_{Z'}^2/g_A)a_0\lambda_t^q r_q$.
	The resulting effective description specifies
	the interactions relevant to the present analysis
	without imposing the additional flavor constraints
	of a particular gauge completion
	\cite{burasstangl2025}.
	
	The scalar leptoquark couples quark and lepton
	fields through its color and weak quantum numbers.
	We consider a field $S_3$ transforming as
	$(\bar{\mathbf 3},\mathbf 3,1/3)$ under the Standard
	Model gauge group
	$SU(3)_c\times SU(2)_L\times U(1)_Y$.
	Thus, it is a color antitriplet and a weak-isospin
	triplet with hypercharge $Y=1/3$. We assume its
	three weak components to have a common mass
	and write their interactions as
	\begin{equation}
		\mathcal L_{S_3}=
		y_{i\mu}\overline{Q_i^c}i\tau_2
		(\tau^aS_3^a)L_\mu+\mathrm{h.c.},
		\label{eq:Slag}
	\end{equation}
	where $Q_i$ and $L_\mu$ denote the left-handed
	quark and muon lepton doublets, respectively,
	$i$ labels the quark generation, and $\tau^a$
	are the Pauli matrices acting on weak-isospin
	indices. The interactions are written in the
	down-quark mass basis, retaining only the muon
	Yukawa column and setting diquark couplings
	to zero.
	
	Integrating out $S_3$ at tree level generates
	a left-handed semileptonic interaction which,
	after a Fierz rearrangement, contributes to
	the axial coefficient $c_A^{q,S_3}$.
	Neutral-meson mixing instead receives its
	leading contribution from box diagrams containing
	two leptoquarks and two leptons. Introducing
	the complex Yukawa products
	$\xi_q=y_{b\mu}y_{q\mu}^*$, the corresponding
	matching coefficients are~\cite{s3cp2021,lqqcd2022}
	$c_A^{q,S_3}=\xi_q/(2M_{S_3}^2)$ and
	$c_{LL}^{q,S_3}=5\xi_q^2/(128\pi^2M_{S_3}^2)$.
	The axial matching gives $\xi_q=2M_{S_3}^2a_0\lambda_t^q r_q$.
	The complex products $\xi_d$ and $\xi_s$
	determine the relevant flavor-changing amplitudes
	and induce correlated kaon transitions; the
	resulting constraints therefore apply to these
	products rather than to individual Yukawa couplings.
	
	To compare the two mediator structures at a common
	mass, we adopt the reference parameters
	$M_{Z'}=M_{S_3}=5\ \mathrm{TeV}$, $g_A=0.5$, and
	$y_{b\mu}=2$ (real).
	The choice of $y_{b\mu}$ also fixes the relation
	between the leptoquark couplings and the induced
	kaon amplitudes.
	
	Quantum chromodynamics (QCD) evolution connects
	the heavy-mediator matching to the meson mixing
	amplitudes. We therefore evolve the new-physics
	contributions to neutral-meson mixing using the
	leading-logarithmic factor~\cite{qcdmixing2000}
	\begin{equation}
		\widehat u(M)=\frac{1}{1.519}
		\left[\frac{\alpha_s(M)}{\alpha_s(m_t)}\right]^{2/7}
		\left[\frac{\alpha_s(m_t)}{\alpha_s(m_b)}\right]^{6/23},
		\label{eq:running}
	\end{equation}
	with two-loop evolution of $\alpha_s$ from
	$\alpha_s(M_Z)=0.1183$ and a normalization factor
	that accounts for the renormalization-group-invariant
	bag-parameter convention.
	
	Combining the matching relations with the QCD
	evolution and normalizing the new contributions
	to the SM mixing amplitude, the coefficient $\kappa_X$
	entering $D_q$ becomes
	$\kappa_{Z'}=\widehat u(M_{Z'})a_0^2M_{Z'}^2/(2g_A^2\widehat c_\SM)$
	for the vector mediator and
	$\kappa_{S_3}=5\widehat u(M_{S_3})a_0^2M_{S_3}^2/(32\pi^2\widehat c_\SM)$
	for the scalar mediator.
	For the reference parameters,
	$\kappa_{Z'}=6.0573$ and
	$\kappa_{S_3}=0.047948$.
	Since the calculation retains leading-order
	matching and leading-logarithmic running,
	we assess higher-order uncertainties through
	variations of the mixing coefficients
	\cite{lqqcd2022}.
	
	To connect these flavor amplitudes to the angular
	observables, we express the semileptonic
	contributions in terms of the Wilson coefficients
	$C_9$ and $C_{10}$, using
	\begin{equation}
		\begin{aligned}
			\mathcal H_{\ell\ell}
			&=-N_b\lambda_t^q(C_9O_9+C_{10}O_{10}),\\
			N_b&=\frac{G_F\alpha_e(4.8\,\mathrm{GeV})}
			{\sqrt{2}\pi}.
		\end{aligned}
	\end{equation}
	The corresponding semileptonic operators are
	$O_9=(\bar q\gamma_\mu\PL b)(\bar\mu\gamma^\mu\mu)$ and
	$O_{10}=(\bar q\gamma_\mu\PL b)(\bar\mu\gamma^\mu\gamma_5\mu)$.
	Here $\alpha_e(4.8\,\mathrm{GeV})$ is the running
	electromagnetic coupling evaluated at the
	indicated scale. The new-physics contributions
	to the Wilson coefficients satisfy
	\begin{equation}
		\delta C_{10}^q=-\frac{a_0r_q}{N_b},
		\qquad
		\delta C_9^q=
		\begin{cases}
			0,&Z',\\
			-\delta C_{10}^q,&S_3.
		\end{cases}
		\label{eq:WC}
	\end{equation}
	For the angular analysis, we denote
	$c\equiv\delta C_{10}^s$.
	These relations establish the connection between
	the complex flavor couplings, neutral-meson
	mixing, and the semileptonic observables used
	in the fit.
	
	\section{Decay, mixing, and time-dependent observables}
	\label{sec:time}
	
	The new-physics contributions to neutral-meson mixing
	modify the oscillation frequencies and CP-violating
	phases according to~\cite{bsmixing2016,latticemixing2016}
	\begin{equation}
		\begin{aligned}
			\Delta m_q&=\frac{2\widehat c_\SM}{3\hbar}
			m_{B_q}f_{B_q}^2\widehat B_q
			|\lambda_t^q|^2|D_q|,\\
			\phi_d^{\rm pred}&=2\beta+\arg D_d
			+\delta_d^{\rm peng},\\
			\phi_s^{\rm pred}&=-2\beta_s+\arg D_s
			+\delta_s^{\rm peng}.
		\end{aligned}
		\label{eq:mix}
	\end{equation}
	Here, $D_q$ is the complex mixing-amplitude ratio
	introduced above. The CKM angles associated with
	$B_d$ and $B_s$ mixing are, respectively,
	$\beta=\arg[-V_{cd}V_{cb}^*/(V_{td}V_{tb}^*)]$ and
	$\beta_s=\arg[-V_{ts}V_{tb}^*/(V_{cs}V_{cb}^*)]$.
	The terms $\delta_q^{\rm peng}$ account for residual
	penguin contributions to the experimentally determined
	mixing phases~\cite{penguins2015}.
	
	For $B_s^0\to\mu^+\mu^-$, the interference between
	the direct decay and neutral-meson mixing amplitudes
	introduces a relative CP-violating phase. In terms
	of the decay-amplitude ratio $P_s$ and the mixing
	ratio $D_s$, we define the phase
	$\theta_s=\arg(P_s^2/D_s)$ and the observables
	$\ADG=\cos\theta_s$ and $\Smm^s=\sin\theta_s$,
	where $\ADG$ governs the decay-width dependence
	of the untagged rate, while $\Smm^s$ determines
	the mixing-induced CP asymmetry. Since the
	interactions considered here modify only the
	axial dimuon amplitude, no independent scalar
	or pseudoscalar decay amplitude is generated.
	Neglecting CP violation in the magnitude of
	neutral-meson mixing, the direct CP-asymmetry
	coefficient therefore vanishes, and the two
	remaining observables satisfy
	$\ADG^2+(\Smm^s)^2=1$~\cite{time2013,cpleptonic2018}.
	
	The modulus of the decay amplitude determines
	the short-distance branching fraction for
	$B_q^0\to\mu^+\mu^-$, which is~\cite{sm2014}
	\begin{equation}
		\BR_q^0=
		\frac{\tau_q f_{B_q}^2m_{B_q}m_\mu^2}
		{8\pi\hbar}
		\sqrt{1-\frac{4m_\mu^2}{m_{B_q}^2}}
		\,|a_0\lambda_t^qP_q|^2,
	\end{equation}
	where $\tau_q$ is the meson lifetime,
	$m_\mu$ is the muon mass, and $P_q=1+r_q$
	contains the new-physics contribution to the
	decay amplitude.
	
	For the $B_s$ system, the nonzero decay-width
	difference modifies the time-integrated branching
	fraction and effective lifetime~\cite{branching2012,lifetime2012}.
	Writing
	$\Gamma_s=(\Gamma_L+\Gamma_H)/2$ and
	$\Delta\Gamma_s=\Gamma_L-\Gamma_H$, where
	$\Gamma_L$ and $\Gamma_H$ are the decay widths
	of the light and heavy mass eigenstates,
	respectively, we introduce
	$y_s=\Delta\Gamma_s/(2\Gamma_s)$ and
	$\tau_s=1/\Gamma_s$. The time-integrated
	branching fraction and effective lifetime
	then read
	\begin{align}
		\overline\BR_s&=
		\BR_s^0\frac{1+\ADG y_s}{1-y_s^2},\\
		\tau_{\mu\mu}^s&=
		\frac{1}{\Gamma_s(1-y_s^2)}
		\frac{1+2\ADG y_s+y_s^2}
		{1+\ADG y_s}.
		\label{eq:taueff}
	\end{align}
	
	The interference between direct decay and
	neutral-meson mixing also produces an oscillatory
	CP asymmetry. When the initial meson flavor is
	identified, this interference remains accessible
	through the tagged decay rates. Defining
	\begin{equation}
		A_{\rm CP}(t)=
		\frac{
			\Gamma(B_s^0(t)\to\mu^+\mu^-)
			-\Gamma(\bar B_s^0(t)\to\mu^+\mu^-)
		}{
			\Gamma(B_s^0(t)\to\mu^+\mu^-)
			+\Gamma(\bar B_s^0(t)\to\mu^+\mu^-)
		},
	\end{equation}
	and neglecting the direct CP contribution, the
	asymmetry takes the form
	\begin{equation}
		A_{\rm CP}(t)=
		\frac{\Smm^s\sin(\Delta m_s t)}
		{\cosh(y_s\Gamma_s t)
			+\ADG\sinh(y_s\Gamma_s t)}.
		\label{eq:tagged}
	\end{equation}
	
	Summing over the initial meson flavor removes
	the oscillatory interference term but retains
	the dependence on the decay-width difference.
	The corresponding normalized untagged decay-time
	distribution is
	\begin{equation}
		\begin{aligned}
			p(t\mid A)=&
			\frac{\Gamma_s(1-y_s^2)}
			{1+A y_s}\,e^{-\Gamma_s t}\\
			&\times\left[
			\cosh(y_s\Gamma_s t)
			+A\sinh(y_s\Gamma_s t)
			\right],
		\end{aligned}
		\label{eq:untagged}
	\end{equation}
	where $A=\ADG$. These ideal distributions
	connect the allowed complex coupling regions
	to temporal predictions before detector
	acceptance, tagging dilution, or time-resolution
	effects are included.
	
	\section{Experimental inputs and theoretical uncertainties}
	\label{sec:data}
	
	We combine rare dimuon decays, neutral-meson mixing, semileptonic and kaon measurements to constrain the complex amplitudes $r_d$ and $r_s$. The leptonic and mixing inputs are summarized in Table~\ref{tab:core}; their values follow the recent LHCb $B_{s,d}^0\to\mu^+\mu^-$ measurements, the HFLAV averages for the 2026 PDG review (including ATLAS measurements), and the independent determination of the $B_d$ mixing phase~\cite{lhcb2026,hflav2026,utfit2025,cosbeta2018}. The CKM parameters are determined from tree-level constraints, independently of the mixing observables under study. Lattice inputs for $B$-meson decay constants and bag parameters are taken from Refs.~\cite{hpqcd2019,flag2024}. The numerical inputs and the kaon matching relations are collected in \ref{app:inputs}. The uncertainties of these quantities enter the fit through fourteen Gaussian-constrained parameters, as described below.
	
	\begin{table}[t]
		\caption{Leptonic and neutral-meson mixing measurements used in the fit. Branching-fraction and lifetime uncertainties include statistical and systematic contributions. The nominal $B_s$ branching fraction assumes $\ADG=1$.}
		\label{tab:core}
		\begin{tabular}{lcc}
			\toprule
			Observable & Value & Uncertainty\\
			\midrule
			$10^{10}\BR_d$ & 1.14 & $^{+0.424}_{-0.393}$\\
			$10^9\overline\BR_s^{\rm nom}$ & 3.25 & $^{+0.291}_{-0.277}$\\
			$\Delta m_d\ [\mathrm{ps}^{-1}]$ & 0.5068 & 0.0019\\
			$\Delta m_s\ [\mathrm{ps}^{-1}]$ & 17.766 & 0.006\\
			$\phi_d^{\rm eff}\ [\mathrm{rad}]$ & 0.77540 & 0.02100\\
			$\phi_s^{c\bar cs}\ [\mathrm{rad}]$ & $-0.053$ & 0.013\\
			$\tau_{\mu\mu}^s\ [\mathrm{ps}]$ & 1.60 & 0.1603
			\\
			\bottomrule
		\end{tabular}
	\end{table}
	
	The LHCb $B_s^0\to\mu^+\mu^-$ branching fraction is quoted assuming $\ADG=1$. To compare it with an arbitrary phase, we use the published acceptance interpolation~\cite{lhcb2026},
	$F_{\rm acc}(A)=1+0.034(1-A)+0.006(1-A)^2$,
	comparing $\overline\BR_s/F_{\rm acc}(\ADG)$ with the nominal measurement. The effect of the acceptance approximation is included in the sensitivity checks.
	
	The initial dataset $\mathcal D_{13}$ comprises the seven observables in Table~\ref{tab:core} and six complementary measurements. These are three lepton-universality ratios $R_K$ (the muon-to-electron partial decay-rate ratios in $B\to K\ell^+\ell^-$), the high-$q^2$ $B^+\to\pi^+\mu^+\mu^-$ rate, the kaon mixing parameter $|\epsilon_K|$, and the branching fraction of $K^+\to\pi^+\nu\bar\nu$. The kaon observables constrain the correlated $s\to d$ coupling induced by the same $S_3$ Yukawa products that govern $b\to d,s$ transitions. We retain the measured $R_K$ correlations and omit $R_{K^*}$ from the primary fit because the cross-analysis covariance is unavailable~\cite{rkstarhigh2026}. The semileptonic and kaon predictions follow Refs.~\cite{flavio2018,wilson2018,kaons2015}.
	
	\subsection{Angular observables}
	\label{sec:angular}
	
	Unlike the leptonic decay, $B^0\to K^{*0}(\to K^+\pi^-)\mu^+\mu^-$ provides access to the interference of different polarization amplitudes. Besides the dilepton invariant mass squared $q^2$, its kinematics is described by the helicity angles $\theta_\ell$ and $\theta_K$ of the dilepton and $K\pi$ systems and the angle $\phi$ between their decay planes, with the flavor-dependent angle convention of Ref.~\cite{angular2026}. After projecting onto the P-wave $K\pi$ state, the differential rate is expanded as
	\begin{equation}
		\frac{\dd^4\Gamma}{\dd q^2\dd\Omega}=
		\frac{9}{32\pi}\sum_i I_i(q^2)f_i(\Omega),
		\label{eq:angular_dist}
	\end{equation}
	where $\dd\Omega=\dd\cos\theta_\ell\,\dd\cos\theta_K\,\dd\phi$. The $f_i$ are the angular basis functions of Ref.~\cite{angular2009,angular2026}; for example, $f_7=\sin 2\theta_K\sin\theta_\ell\sin\phi$, $f_8=\sin 2\theta_K\sin 2\theta_\ell\sin\phi$, and $f_9=\sin^2\theta_K\sin^2\theta_\ell\sin 2\phi$. The coefficients $I_i$ are bilinear in the transversity amplitudes $A_\lambda^{L,R}$, where $\lambda=0,\parallel,\perp$ denotes the longitudinal or either of the two transverse polarization combinations, and $L,R$ refer to the chirality of the lepton current.
	
	The operators $O_9$ and $O_{10}$ in Sec.~\ref{sec:models} contribute through the combinations $C_9\mp C_{10}$, since the vector and axial currents are respectively the sum and difference of the chiral currents. The dependence on the additional Wilson coefficients can thus be written as
	\begin{equation}
		A_\lambda^{L,R}=A_{\lambda,\SM}^{L,R}
		+F_\lambda(q^2)(\delta C_9^s\mp\delta C_{10}^s),
		\label{eq:transversity_np}
	\end{equation}
	with the upper (lower) sign for $L$ ($R$). The functions $F_\lambda$ contain the local hadronic form factors and the normalization and kinematic factors of the corresponding transversity amplitude. The SM amplitudes also include the electromagnetic dipole and nonlocal hadronic contributions~\cite{angular2009,lowrecoil2010,charm2010,dispersion2018}. Equation~\eqref{eq:WC} therefore gives shifts proportional to $(-c,+c)$ in the left- and right-handed amplitudes for $Z'$, and to $(-2c,0)$ for $S_3$. The two interactions modify different combinations of polarization amplitudes even when their contributions to the leptonic decay coincide.
	
	In the massless-lepton limit, the three imaginary interference coefficients of particular interest have the structure~\cite{angular2009}
	\begin{align}
		I_7&=\sqrt{2}\,\Im\big[A_0^LA_\parallel^{L*}
		-A_0^RA_\parallel^{R*}\big],\notag\\
		I_8&=\frac{1}{\sqrt{2}}\Im\big[A_0^LA_\perp^{L*}
		+A_0^RA_\perp^{R*}\big],\notag\\
		I_9&=\Im\big[A_\parallel^{L*}A_\perp^L
		+A_\parallel^{R*}A_\perp^R\big].
		\label{eq:imag_interference}
	\end{align}
	Here the transversity phases are chosen as in Ref.~\cite{angular2009}; the coefficients are expressed in the experimental angular convention when comparing with data. Denote by $\bar I_i$ the coefficients of the CP-conjugate process in CP-related angular coordinates, with weak phases reversed and strong phases unchanged. Taking the unbarred rate to refer to the $b$-quark decay, the CP-averaged observables and CP asymmetries in a bin $b$ are
	\begin{equation}
		\begin{aligned}
			S_{i,b}&=\frac{\int_b\dd q^2\,(I_i+\bar I_i)}{G_b},\\
			A_{i,b}&=\frac{\int_b\dd q^2\,(I_i-\bar I_i)}{G_b},\\
			G_b&=\int_b\dd q^2\left(\frac{\dd\Gamma}{\dd q^2}
			+\frac{\dd\bar\Gamma}{\dd q^2}\right),
		\end{aligned}
		\label{eq:angular_obs}
	\end{equation}
	where $G_b$ is the combined P-wave partial rate. In this convention $A_{\rm FB}=3S_6^s/4$ for the selected model. In particular, the imaginary part of an interference product with relative strong and weak phases $\Delta\delta$ and $\Delta\varphi$ is proportional to $\sin(\Delta\delta+\Delta\varphi)$. Subtracting its CP conjugate gives a term proportional to $2\cos\Delta\delta\sin\Delta\varphi$. The asymmetries $A_{7,8,9}$ can therefore probe relative weak phases even when strong-phase differences are small. The CP-averaged observables retain complementary information on the magnitudes and interference of the same amplitudes.
	
	We use the LHCb CP-floating configuration (iv),
	in which the CP asymmetries are allowed to vary
	in the angular fit, together with its published
	experimental covariance~\cite{angular2026,angular2026data}.
	Configuration (iv) uses the massless-lepton angular model. We retain only the P-wave observables associated with the $K^{*0}$ resonance.
	In each dilepton-mass bin $b$, we retain the observable vector
	$\bm a_b=(S_{2c},S_3,S_4,S_5,A_{\rm FB},\allowbreak
	S_7,S_8,S_9,A_7,A_8,A_9)_b$,
	where $S_{2c}=-F_L$ determines the longitudinal
	$K^{*0}$ polarization fraction in the massless-lepton
	limit. The observable $S_3$ probes the relative
	transverse polarizations, while $S_4$ and $S_5$
	describe longitudinal--transverse interference.
	The forward--backward asymmetry $A_{\rm FB}$
	characterizes the angular preference of the muons,
	whereas $S_7$, $S_8$, and $S_9$ probe imaginary
	interference terms between transversity amplitudes.
	The corresponding CP asymmetries $A_{7,8,9}$
	provide sensitivity to their weak phases.
	
	We select the five nonoverlapping $q^2$ intervals
	$[1.1,2.5]$, $[2.5,4]$, $[4,6]$, $[15,17]$,
	and $[17,19]\ \mathrm{GeV}^2$, avoiding the
	photon-pole and resonance regions. Starting from
	the 13 nonangular flavor observables in
	$\mathcal D_{13}$, adding the three CP asymmetries
	in each bin gives $\mathcal D_{28}$, while including
	all eleven selected angular observables per bin
	defines $\mathcal D_{68}$.

	The $55\times55$ experimental covariance is reconstructed from the published total correlation matrix $R^{\rm total}$ and the bias-corrected uncertainties $\sigma_i=(\sigma_i^++|\sigma_i^-|)/2$, so that $(V_{\rm ang})_{ij}=R_{ij}^{\rm total}\sigma_i\sigma_j$.
	It includes the experimental statistical, systematic and Neyman terms without double counting. The selected principal submatrix preserves the published correlations. The CP-asymmetry inputs and their within-bin covariance elements are collected in \ref{app:inputs}.
	
	The linear dependence in Eq.~\eqref{eq:transversity_np} implies that every integrated angular numerator and the normalization $G_b$ are linear combinations of $1$, $\Re c$, $\Im c$, and $|c|^2$ at fixed hadronic inputs. The observables are obtained by taking the ratios in Eq.~\eqref{eq:angular_obs}, not by averaging the normalized quantities across a bin. For real local form factors, the opposite axial shifts in the two chiral amplitudes cancel the weak-phase-dependent contribution to the unnormalized $A_8$ and $A_9$ numerators for $Z'$ at the retained order. The vector shift in the $S_3$ scenario also modifies these numerators; the normalization can change for either mediator. Thus, the angular measurements constrain the two interaction structures differently.
	
	The hadronic inputs affect several angular coefficients simultaneously, so their uncertainties cannot be assigned independently to each observable. We propagate the form-factor~\cite{bsz2016} and nonlocal-amplitude uncertainties through a common ensemble of hadronic inputs at every value of the complex coupling. This construction retains correlations between observables and $q^2$ bins, as well as their dependence on the new-physics coefficients. The numerical implementation, including the treatment of CKM conjugation, is described in \ref{app:checks}.
	
	\subsection{Statistical inference and calibration}
	\label{sec:inference}
	
	The aim of the fit is to determine which values of $r_d$ and $r_s$ can reproduce the measured observables once experimental and theoretical uncertainties are taken into account. Let $\bm d$ be the vector of measured central values and $\bm\mu(\bm r,\bm\nu)$ the corresponding predictions, where $\bm r=(\Re r_d,\Im r_d,\Re r_s,\Im r_s)$. The fourteen components of $\bm\nu$ are standardized Gaussian variables describing shifts of the constrained flavor, hadronic and short-distance inputs, with correlations included in the mapping to the physical inputs. Correlated theory errors are represented by an additional shift $J\bm\eta$, where $J$ is constructed from the common ensembles and the components of $\bm\eta$ have unit Gaussian constraints. The residual is $\bm e=\bm\mu+J\bm\eta-\bm d$, and the fit minimizes $\chi^2=\bm e^TV_{\rm exp}^{-1}\bm e+\bm\nu^T\bm\nu+\bm\eta^T\bm\eta$. The inverse experimental covariance $V_{\rm exp}^{-1}$ weights the differences between predictions and measurements according to their uncertainties and correlations, while the last two terms penalize shifts of the constrained inputs.
	
	For a Gaussian correlated block, minimizing with respect to $\bm\eta$ at fixed $\bm r,\bm\nu$ is equivalent to using the covariance $V_{\rm exp}+JJ^T$ in the residual term with $\bm\eta=0$. This minimization is called profiling. Since the theory shifts are minimized over rather than integrated out, no logarithmic determinant term is introduced. Asymmetric experimental errors are treated using the uncertainty on the side of the measured central value selected by the residual.
	
	To constrain a physical quantity $h$, such as $|\xi_s|$ or $\theta_s$, we fix it at a value $h_0$ and minimize over all remaining parameters. The deterioration in the description of the data relative to the unrestricted minimum is measured by~\cite{cowan2011}
	\begin{equation}
		q_h(h_0)=\min_{h=h_0}\chi^2-\min\chi^2.
		\label{eq:q}
	\end{equation}
	The same construction applies to a pair of quantities, such as $(\Re r_s,\Im r_s)$. Minimizations from multiple starting points account for disconnected solutions. The physical coupling moduli include the CKM and short-distance normalization factors, and both complex solutions of Eq.~\eqref{eq:quadratic} relating $r_s$ to $\theta_s$ are retained.
	
	For regular likelihoods in the large-sample limit, the 95\% thresholds for one and two constrained quantities are $3.8415$ and $5.9915$, respectively~\cite{wilks1938,cowan2011}. This approximation may fail here: the coupling modulus cannot be negative and its phase is undefined at zero coupling. We therefore determine the cutoffs numerically by generating artificial datasets from specified parameter values and repeating the fit. For each generating point, the 95th percentile of the resulting $q_h$ distribution estimates the cutoff that would retain the generating value in 95\% of repeated measurements. We use the larger of the upper statistical estimates of this percentile at the profiled SM point and at the best-fit point for each mediator. Independent pseudoexperiments at selected boundary points then test how often the corresponding generating values are retained. The calibration and validation refer to the tested generating points and are performed separately for each reported profile~\cite{chernoff1954,selfliang1987,feldman1998}. The calibration prescription and associated numerical results are collected in \ref{app:statistics}.
	
	\section{Angular constraints and time-dependent predictions}
	\label{sec:results}
	
	The minimum $\chi^2$ values for $\mathcal D_{13}$, $\mathcal D_{28}$, and $\mathcal D_{68}$ (Table~\ref{tab:fits}) are comparable for the two mediators, although their preferred complex amplitudes respond differently to the angular constraints.
	
	The strongest effect occurs in the $S_3$ scenario. At the adopted masses and couplings, $\kappa_{S_3}=0.047948$ is much smaller than $\kappa_{Z'}=6.0573$, reflecting the loop suppression of the leptoquark contribution to neutral-meson mixing. The same complex $r_s$ can therefore produce a much smaller departure from SM mixing for $S_3$. Without angular constraints, this permits a large imaginary $b\to s$ amplitude, corresponding to a best-fit value $\Smm^s\simeq0.95$. Including the CP asymmetries $A_{7,8,9}$ reduces this value to approximately $0.145$, while the complete selected P-wave dataset gives $\Smm^s\simeq0.103$. By comparison, the preferred $Z'$ value changes only from approximately $0.053$ to $0.044$.
	
	The restriction of the scalar phase also reduces the allowed coupling-modulus projection. At the same nominal one-dimensional threshold, the upper limit on
$|\xi_s|$ falls from $0.16239$ to $0.02749$, a reduction of about
$83\%$, while the change for $Z'$ is smaller. Table~\ref{tab:sequential} summarizes this comparison at a common nominal threshold, which isolates the impact of the additional measurements; the final coupling constraints are obtained from the calibrated profiles.
	
	\begin{table}[t]
  \caption{Upper coupling-modulus projections and the $S_3$ temporal
    asymmetry obtained with the same nominal one-parameter threshold
    $\Delta\chi^2=3.8415$ for each data selection.}
  \label{tab:sequential}
  \centering\small
  \begin{tabular}{lccc}
  \toprule
  Dataset & max $|g_{sb}|$ & max $|\xi_s|$ & $\Smm^s$ ($S_3$) \\
  \midrule
  $\mathcal D_{13}$ & 0.01671 & 0.16239 & $[-0.986,\,1.000]$ \\
  $\mathcal D_{28}$ & 0.01676 & 0.03755 & $[-0.127,\,0.426]$ \\
  $\mathcal D_{68}$ & 0.01547 & 0.02749 & $[-0.106,\,0.311]$ \\
  \bottomrule
  \end{tabular}
\end{table}
	
\begin{figure*}[!t]
		\centering
		\includegraphics[width=.98\textwidth]{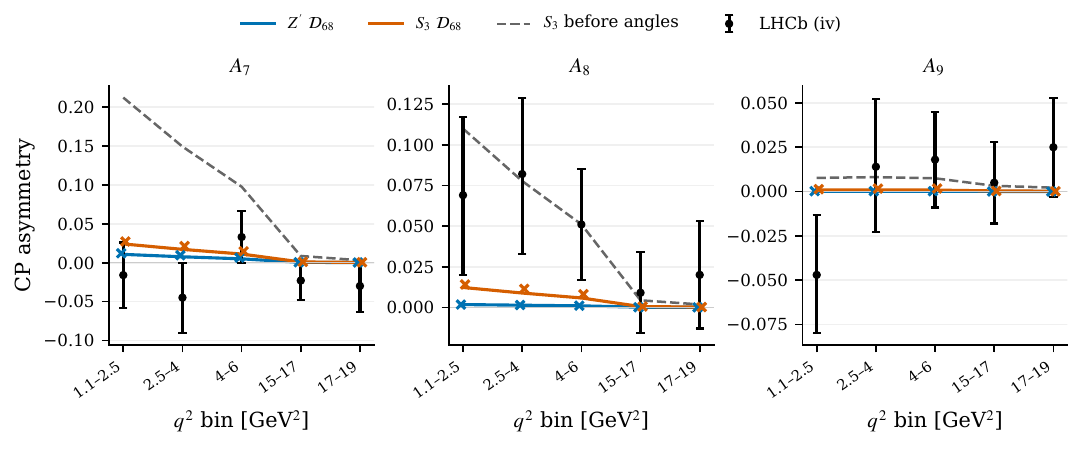}
		\caption{CP asymmetries $A_7$, $A_8$, and $A_9$ in the selected dilepton-mass bins, compared with the fitted $Z'$ and $S_3$ predictions. Experimental points include the published asymmetric uncertainties. Curves represent central-input predictions, while smaller markers include profiled theoretical shifts. The pre-angular $S_3$ solution illustrates the large-phase configuration disfavored by the angular measurements. Experimental correlations are included in the fit.}
		\label{fig:angular}
	\end{figure*}

	The restriction of large scalar phases is also visible in Fig.~\ref{fig:angular}, which compares the measured CP asymmetries with the predictions of the two fitted scenarios. The pre-angular $S_3$ solution illustrates the large-phase configuration disfavored by the angular measurements, whereas the refitted predictions remain compatible with the measured asymmetries within the reconstructed experimental and theoretical uncertainties.
	
	The transformation $r_s\mapsto-2-r_s$ sends $P_s$ to $-P_s$ and
leaves the short-distance dimuon rate unchanged. In the $S_3$ model,
it acts differently on the angular amplitudes. With the conventions of
Eq.~\eqref{eq:transversity_np}, write
\begin{equation}
  \begin{aligned}
    A_\lambda^L(r_s) &= H_\lambda+X_\lambda(r_s),\quad
    X_\lambda(r_s)=\frac{2a_0}{N_b}F_\lambda(1+r_s),\\
    H_\lambda &= A_{\lambda,\SM}^L-\frac{2a_0}{N_b}F_\lambda .
  \end{aligned}
  \label{eq:branch_split_transversity}
\end{equation}
The reflection reverses $X_\lambda$ and leaves $H_\lambda$ and
$A_\lambda^R$ fixed. If $H_\lambda$ vanished for every helicity, the
reflection would simply change the common sign of the left-handed
amplitudes, leaving their bilinears invariant. In the physical amplitude,
$H_\lambda$ contains the electromagnetic dipole and nonlocal hadronic
terms. The local term proportional to $F_\lambda$ carries the coefficient
$C_9^{\SM}-C_{10}^{\SM}-2a_0/N_b$.
For example, subtracting the reflected value of
$\operatorname{Re}(A_i^LA_j^{L*})$ from its ordinary value gives
$2\operatorname{Re}(H_iX_j^*+X_iH_j^*)$.
$S_5$ probes the longitudinal--perpendicular pair and $A_{\rm FB}$
the parallel--perpendicular pair. At the central hadronic inputs,
the dipole term gives the largest contribution to the change of both
unnormalized angular moments in each of the three low-$q^2$ bins.
The fit uses the bin-integrated normalized observables.

\begin{figure*}[!t]
  \centering
  \includegraphics[width=.98\textwidth]{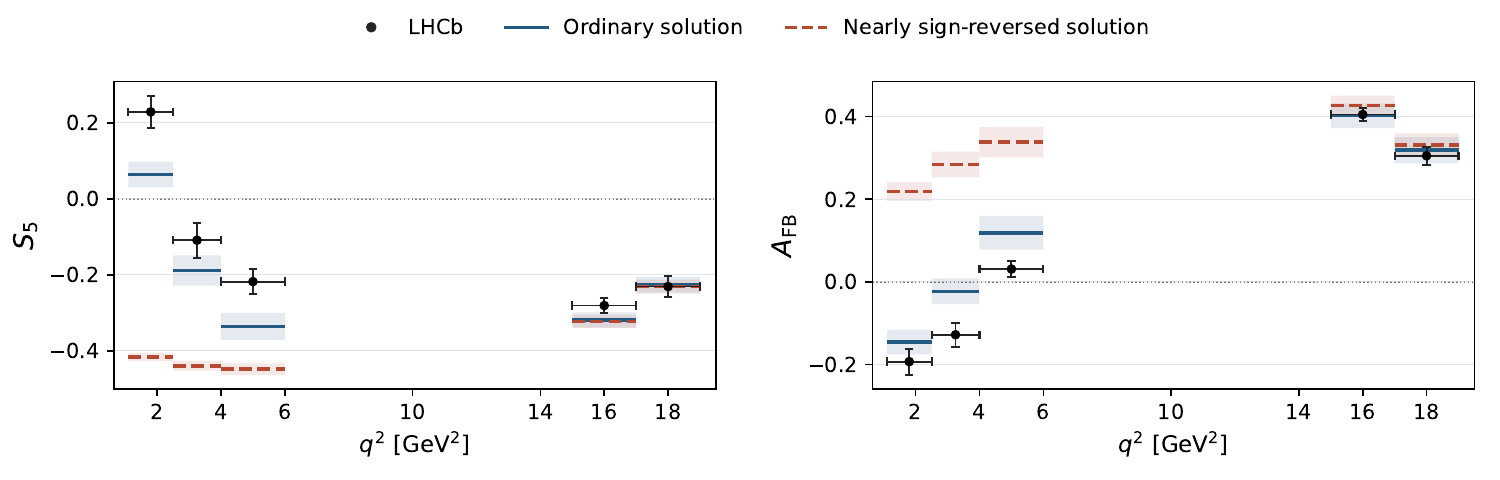}
  \caption{The two $S_3$ solutions of the $\mathcal D_{28}$ fit evaluated
    in each selected bin for $S_5$ and $A_{\rm FB}$. The black markers
    show the LHCb measurements and their total uncertainties. Shaded
    bands give the marginal hadronic standard deviations at the two
    fitted amplitude points. The correlated errors are used in the
    fit.}
  \label{fig:branch_s5_afb}
\end{figure*}

The fit to $\mathcal D_{28}$ still has a nearly sign-reversed solution
at $r_s\simeq-1.91+0.10i$, just inside the nominal two-dimensional
contour. Figure~\ref{fig:branch_s5_afb} shows the $S_5$ and $A_{\rm FB}$
predictions from the two $\mathcal D_{28}$ minima, before these
observables enter the fit. The separation is most visible in the
low-mass bins. We quantify it by
\begin{equation}
  \Delta\chi^2_{\rm sec} =
  \min_{\operatorname{Re}r_s\leq-1}\chi^2
  -\min_{\operatorname{Re}r_s\geq-1}\chi^2 ,
  \label{eq:branch_sector_gap}
\end{equation}
profiling the remaining flavor and nuisance parameters on each side.
The gap in $\mathcal D_{28}$ is $5.967$. With the full
angular sample, the minimum in the inverted sector lies on the
boundary at $r_s\simeq-1+0.965i$. The local minimum near
$r_s\simeq-1.90+0.13i$ has a larger separation of $573.8$.
Of the individual CP-averaged observables, $S_5$ and $A_{\rm FB}$
produce the largest increases in $\Delta\chi^2_{\rm sec}$.

Table~\ref{tab:branch_recoil} shows how the low- and high-$q^2$
measurements contribute when fitted separately. The low-$q^2$
sample supplies most of the separation. This pattern persists when
the residual nonlocal amplitudes are given a different $q^2$
dependence with the same average prior variance, as described in
Appendix~\ref{app:checks}.

\begin{table}[t]
  \caption{Profiled $S_3$ sector separations for two prescriptions
    for the residual nonlocal amplitudes. The low- and high-$q^2$
    selections each add the CP-asymmetric and CP-averaged $P$-wave
    observables in their stated bins to $\mathcal D_{13}$.}
  \label{tab:branch_recoil}
  \centering\small
  \begin{tabular}{lrr}
    \toprule
    Angular input & Reference & Alternative \\
    \midrule
    $\mathcal D_{28}$ with $S_5$ and $A_{\rm FB}$ & 340.5 & 342.3 \\
    $q^2\in[1.1,6]\,\mathrm{GeV}^2$ & 375.6 & 384.9 \\
    $q^2\in[15,19]\,\mathrm{GeV}^2$ & 7.9 & 7.5 \\
    Full $\mathcal D_{68}$ & 423.4 & 437.1 \\
    \bottomrule
  \end{tabular}
\end{table}

\begin{figure*}[!t]
		\centering
		\includegraphics[width=.98\textwidth]{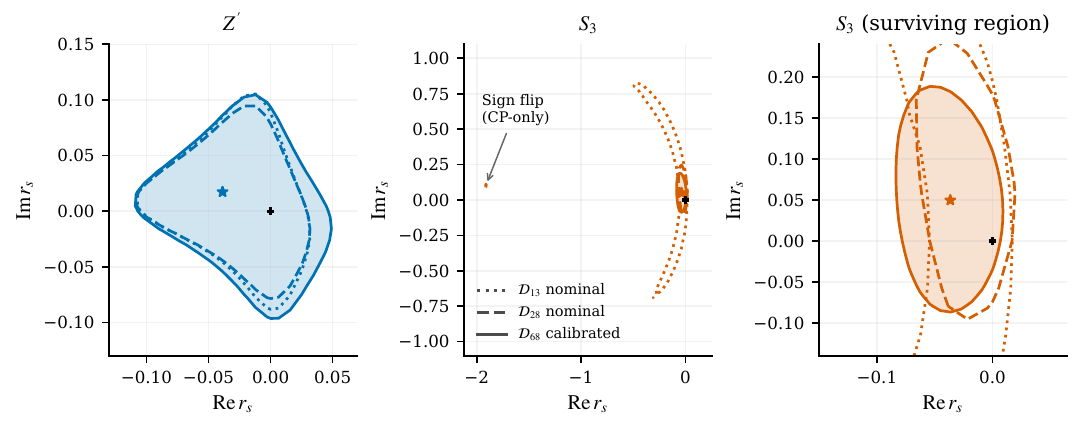}
		\caption{Allowed regions in the complex $r_s$ plane
			for the $Z'$ (blue, left panel) and $S_3$ (orange,
			central and right panels) scenarios.
			Dotted and dashed contours correspond to the nominal
			two-dimensional profiles for $\mathcal D_{13}$ and
			$\mathcal D_{28}$, respectively, at
			$\Delta\chi^2=5.9915$, whereas solid contours enclose
			the regions obtained from the complete $\mathcal D_{68}$
			fit using the separately calibrated thresholds.
			The corresponding allowed regions are shaded.
			Stars indicate the $\mathcal D_{68}$ best-fit points,
			while black crosses mark $r_s=0$.
			The central panel also displays the approximately
			sign-inverted scalar solution retained when only the
			CP-asymmetry measurements are added
			($\mathcal D_{28}$), which is excluded once the
			complete angular information is included.
			The right panel enlarges the surviving $S_3$ region.
			The remaining flavor amplitudes and theoretical
			parameters are profiled throughout.}
		\label{fig:regions}
	\end{figure*}

	The resulting allowed regions are illustrated in Fig.~\ref{fig:regions}. The final complex-amplitude regions are obtained by profiling over the remaining flavor and theoretical parameters with the calibrated two-dimensional threshold.
	
	\begin{figure*}[!t]
		\centering
		\includegraphics[width=.98\textwidth]{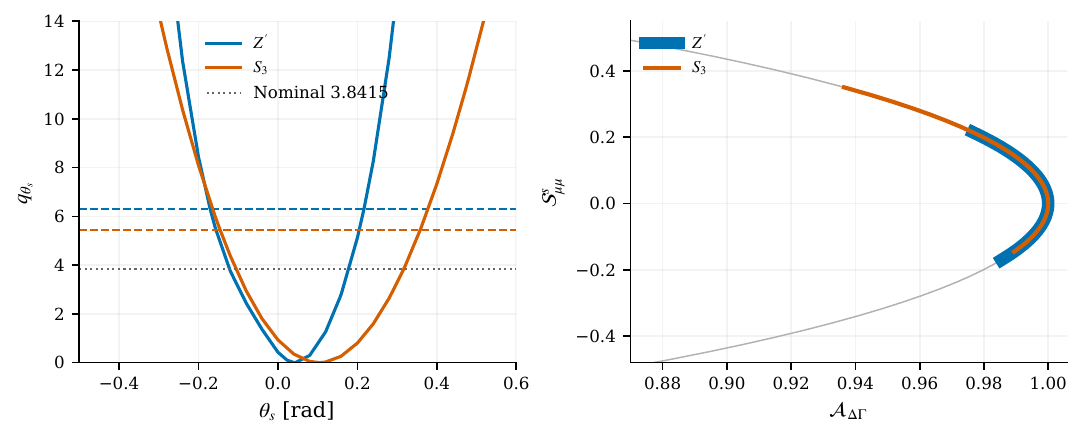}
		\caption{Constraints on the phase $\theta_s$ and their
			implications for the temporal observables.
			Left panel: profile statistics $q_{\theta_s}$ for the
			$Z'$ (blue) and $S_3$ (orange) scenarios obtained from
			the complete angular fit. The solid curves represent
			the phase profiles, while the horizontal dashed lines
			indicate the corresponding calibrated cutoffs.
			The gray dotted line marks the nominal one-parameter
			threshold $q_{\theta_s}=3.8415$.
			Right panel: the accepted phase intervals mapped onto
			the $(\ADG,\Smm^s)$ plane. The colored segments
			represent the allowed regions for each mediator,
			whereas the thin gray curve shows the relation
			$\ADG^2+(\Smm^s)^2=1$.
			Both observables are determined by the same phase
			and are therefore correlated.}
		\label{fig:phase}
	\end{figure*}
	
	The best-fit configurations (Table~\ref{tab:best}) retain a larger imaginary $b\to d$ amplitude for the scalar mediator, despite the restriction of its $b\to s$ phase.
	
	The separately calibrated profiles give the physical coupling constraints and temporal predictions summarized in Table~\ref{tab:bounds}, with coupling-modulus intervals that include zero for both scenarios.
	
	\begin{table}[t]
  \caption{Calibrated individual projections from $\mathcal D_{68}$.
    Both temporal observables follow from the same allowed phase set.}
  \label{tab:bounds}
  \centering\small
  \begin{tabular}{lcc}
  \toprule
  Quantity & $Z'$ & $S_3$ \\
  \midrule
  $|g_{db}|$ or $|\xi_d|$ & $[0,\,0.00757]$ & $[0,\,0.06267]$ \\
  $|g_{sb}|$ or $|\xi_s|$ & $[0,\,0.01900]$ & $[0,\,0.03388]$ \\
  $\theta_s\ [\mathrm{rad}]$ & $[-0.171,\,0.217]$ & $[-0.146,\,0.358]$ \\
  $\Smm^s$ & $[-0.170,\,0.215]$ & $[-0.145,\,0.351]$ \\
  $\ADG$ & $[0.977,\,1.000]$ & $[0.936,\,1.000]$ \\
  \bottomrule
  \end{tabular}
\end{table}
	
	The phase profiles give $\ADG\gtrsim0.977$ for $Z'$ and
$\ADG\gtrsim0.936$ for $S_3$. Both untagged distributions thus
stay close to the heavy-eigenstate limit.
	
	Figure~\ref{fig:phase} shows the calibrated phase profiles and their mapping onto the temporal observables. Since $\Smm^s$ and $\ADG$ depend on the same phase $\theta_s$, their allowed values lie on a common curve rather than filling the Cartesian product of the intervals in Table~\ref{tab:bounds}.
	
	The time-dependent curves in Fig.~\ref{fig:time} illustrate the different phase sensitivity of tagged and untagged rates. Near $\theta_s=0$, the tagged asymmetry is linear in the phase since $\Smm^s\simeq\theta_s$, whereas the untagged rate changes only at quadratic order since $\ADG\simeq1-\theta_s^2/2$. This explains the greater intrinsic phase sensitivity of the tagged observable.
	
	At the central decay widths, the branching-fraction acceptance
correction is at most about $0.22\%$ over the accepted phase
regions. The corresponding effective lifetimes are
$[1.6088,\,1.6107]\,\mathrm{ps}$ for $Z'$ and $[1.6056,\,1.6107]\,\mathrm{ps}$ for $S_3$. These shifts are smaller than the uncertainty associated with the decay widths and considerably below the present experimental lifetime uncertainty. Although $S_3$ allows a wider range of tagged asymmetries, the two predictions overlap substantially.
	
	To assess the effect of the updated dimuon measurements independently of the angular data, we repeat the complete angular fit using the previous Run~1--2 branching fractions and effective lifetime \cite{lhcb2021}, while keeping all other inputs unchanged. At the same nominal profile threshold, the upper scalar coupling projection changes only from $|\xi_s|=0.02733$ to $0.02748$, while the vector projection changes from $|g_{sb}|=0.01635$ to $0.01547$. The corresponding temporal-phase ranges also change only slightly. The substantial reduction of the previously allowed scalar region is therefore driven primarily by the angular measurements rather than by the updated dimuon data alone. This comparison uses the same archived angular response for the two dimuon data sets.
	
	Finally, the comparison between the two mediators depends on their masses and coupling structures. Equal values of $r_d$, $r_s$, and $\kappa_X$ produce identical leptonic and mixing observables at the retained order. Neglecting the weak mass dependence of the running, the corresponding masses satisfy
	$M_{S_3}/M_{Z'}\simeq4\pi/(\sqrt{5}|g_A|)$.
	Including the adopted running, a $5$ TeV vector mediator corresponds to a scalar mass near $58$ TeV under this restricted matching condition. This degeneracy does not extend to the complete fit, since the two interactions generate different semileptonic operator relations and kaon contributions.
	
\begin{figure*}[!t]
  \centering
  \includegraphics[width=.98\textwidth]{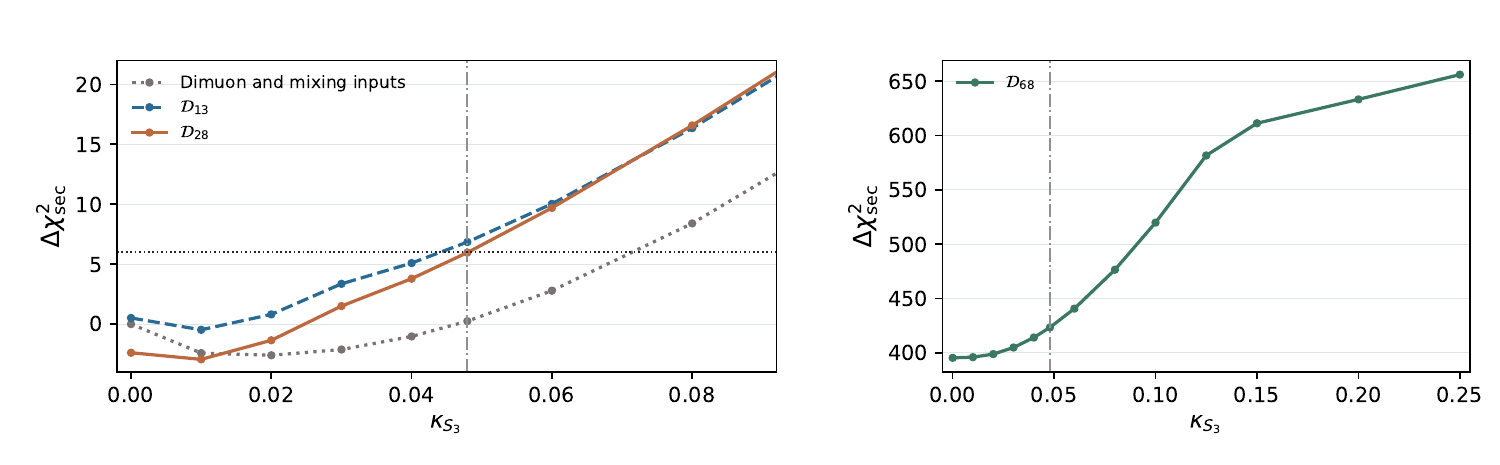}
  \caption{Profiled sector separation in the $S_3$ scenario as the
    coefficient of the new contribution to $B$ mixing is varied.
    The vertical line marks the reference value and the horizontal
    dotted line in the left panel is the nominal two-dimensional
    threshold. Negative values indicate a lower minimum on the
    inverted side. The two panels use different vertical scales.}
  \label{fig:branch_kappa}
\end{figure*}

	We vary $\kappa_{S_3}$ in both $B_d$ and $B_s$ mixing while keeping
the rest of the matching fixed. At $r_s=-2$ the shift in the
$B_s$ mixing amplitude ratio is $D_s-1=4\kappa_{S_3}$.
Figure~\ref{fig:branch_kappa} shows how the sector separation responds
to this change after profiling the two amplitudes. Using the nominal
criterion $\Delta\chi^2_{\rm sec}=5.9915$ for comparison, the separation
reaches this threshold at $\kappa_{S_3}\simeq0.0721$ with
the dimuon and mixing inputs alone. The crossing is at $0.0443$
for $\mathcal D_{13}$ and $0.0480$ for $\mathcal D_{28}$.
The reference value $0.047948$ thus lies close to the last crossing.
Once the full angular sample is included, the separation is already
about $395$ at $\kappa_{S_3}=0$ and remains well above the
nominal threshold across the scan.

	\subsection{Statistical validation and scope}
	\label{sec:robustness}
	
	The fitted amplitudes, the SM and the simulated data sets all use
the same weak-phase treatment described in Appendix~\ref{app:checks}.
At the eight tested boundary points, independent pseudoexperiments
retain the generating values in $95.6\%$--$99.8\%$ of the
profiles (Table~\ref{tab:validation}). The resulting conditional
tail probabilities for the improvement over the SM are $0.158$
for $Z'$ and $0.042$ for $S_3$.

	\section{Conclusions}
	\label{sec:conc}
	
	Combining the November 2026 LHCb rare dimuon measurements with neutral-meson mixing, semileptonic and kaon constraints and the recent $B^0\to K^{*0}\mu^+\mu^-$ angular analysis, we have traced how the angular information resolves complex flavor-amplitude degeneracies for a heavy $Z'$ and a scalar triplet leptoquark $S_3$. The suppression of the scalar mixing contribution allows large weak phases and an approximately sign-inverted leptonic decay amplitude. Angular CP asymmetries restrict the former, whereas the CP-averaged observables resolve the latter. The separate angular fits identify $S_5$ and $A_{\rm FB}$ as the leading probes of the sign-inverted branch through their sensitivity to vector--axial interference. The full angular sample separates the two $S_3$ amplitude sectors by $\Delta\chi^2_{\rm sec}\simeq423$ at the reference mixing coefficient and by about $395$ even at $\kappa_{S_3}=0$. At a common nominal one-dimensional threshold, the upper projection on the strange-sector scalar coupling decreases by approximately $83\%$.
	
	The removal of these amplitude solutions also determines the range of time-dependent CP effects accessible in $B_s^0\to\mu^+\mu^-$. The calibrated intervals for $\Smm^s$ are $[-0.170,\,0.215]$ for $Z'$ and
$[-0.145,\,0.351]$ for $S_3$. Both interactions therefore admit substantially overlapping ranges of tagged CP asymmetries, while their untagged decay-time distributions remain close to the heavy-eigenstate limit. The combined angular and temporal observables provide complementary information: the former distinguish the complex-amplitude branches, and the latter probe the surviving relative phase between decay and mixing.
	
	\section*{Data and code availability}
	The data and code are not publicly available. They are available from the authors upon reasonable request.
	
	\section*{Acknowledgments}
	A. C. and G. P. acknowledge partial financial support from the Italian
	Ministry of Education, University and Research (MIUR).
	The authors acknowledge support from the Istituto Nazionale di Fisica
	Nucleare (INFN) through the Iniziativa Specifica QGSKY.
	
	\FloatBarrier
	\appendix
	\section{Experimental and theoretical input details}
	\label{app:inputs}
	
	For independent CKM information we use the tree-level determination~\cite{utfit2025}
	\begin{align}
		s_{12}&=0.2251\pm0.0006,&|V_{cb}|&=0.04118\pm0.00076,\nonumber\\
		|V_{ub}|&=0.00382\pm0.00034,&\gamma&=(65.7\pm2.5)^\circ.
		\label{eq:CKM}
	\end{align}
	The lattice normalization is~\cite{hpqcd2019,flag2024}
	\begin{align}
		f_{B_s}&=(0.2303\pm0.0013)\ \mathrm{GeV},\nonumber\\
		f_{B_s}/f_{B_d}&=1.209\pm0.005,\nonumber\\
		\widehat B_s&=1.232\pm0.053,\nonumber\\
		\widehat B_s/\widehat B_d&=1.008\pm0.025.
		\label{eq:lattice}
	\end{align}
	These inputs are treated as independent Gaussian constraints, with bag-parameter correlations varied in the sensitivity study. We use $\Gamma_s=0.6600\pm0.0024\ \mathrm{ps}^{-1}$, $\Delta\Gamma_s=0.0783\pm0.0035\ \mathrm{ps}^{-1}$ with correlation $-0.09$, and Gaussian residual penguin-phase shifts of $0\pm0.005$ rad. The $B_d$ lifetime is fixed to $\tau_d=1.5113$ ps.
	
	The complementary lepton-universality ratios are
	$R_K^{[1.1,6]}=0.949^{+0.048}_{-0.047}$,
	$R_K^{[0.1,1.1]}=0.994^{+0.095}_{-0.086}$, and
	$R_K^{q^2>14.3}=1.08^{+0.117}_{-0.098}$~\cite{lfu2023,rkhigh2025}.
	The high-$q^2$ $B^+\to\pi^+\mu^+\mu^-$ input is
	$10^9\langle\dd\BR_{\pi\mu\mu}/\dd q^2\rangle_{[15,22]}$
	$=0.47^{+0.12}_{-0.10}\pm0.01\ \mathrm{GeV}^{-2}$~\cite{pimumu2015}.
	The published correlation between the first two $R_K$ measurements is retained, while missing correlations with the high-$q^2$ measurement are set to zero. An independent $1\%$ relative QED prediction uncertainty is used for each $R_K$~\cite{lfu2023,rkqed2016}.
	
	The $S_3$ Yukawa products induce the correlated kaon coupling
	$\zeta_{sd}=y_{s\mu}y_{d\mu}^*=\xi_s^*\xi_d/y_{b\mu}^2$ and
	the mixing coefficient $c_{LL}^{K,S_3}=5\zeta_{sd}^2/(128\pi^2M_{S_3}^2)$.
	The muon-neutrino contribution is
	$\delta(\lambda_t^K X)_\mu=-s_W^2\zeta_{sd}^*/(4N_KM_{S_3}^2)$,
	where $N_K=G_F\alpha_e(2\,\mathrm{GeV})/(\sqrt{2}\pi)$,
	$s_W=\sin\theta_W$, with $\theta_W$ the weak mixing angle,
	$\lambda_t^K=V_{ts}^*V_{td}$, and $X$ denotes the
	short-distance function entering the $s\to d\nu\bar\nu$ amplitude.
	They are constrained by $|\epsilon_K|=(2.228\pm0.011)\times10^{-3}$ and
	$\BR(K^+\to\pi^+\nu\bar\nu)=(9.6^{+1.9}_{-1.8})\times10^{-11}$~\cite{pdg2024,na622026}.
	The asymmetries $A_7,A_8,A_9$ in the five selected bins are given in Table~\ref{tab:cp-data}, and their within-bin covariance elements in Table~\ref{tab:cp-cov}; the complete $55\times55$ selected matrix, including the other P-wave observables, is constructed directly from the published HEPData input~\cite{angular2026data}.
	
	\begin{table}[t]
		\caption{Bias-corrected CP asymmetries from the LHCb CP-floating angular analysis in the selected dilepton-mass bins. Bin boundaries are in $\mathrm{GeV}^2$.}
		\label{tab:cp-data}
		\centering\small
		\begin{tabular}{lccc}
			\toprule
			$q^2$ bin & $A_7$ & $A_8$ & $A_9$\\
			\midrule
			$1.1$--$2.5$ &
			$-0.016^{+0.042}_{-0.042}$ &
			$0.069^{+0.048}_{-0.049}$ &
			$-0.047^{+0.034}_{-0.033}$\\
			$2.5$--$4$ &
			$-0.045^{+0.045}_{-0.046}$ &
			$0.082^{+0.047}_{-0.049}$ &
			$0.014^{+0.038}_{-0.037}$\\
			$4$--$6$ &
			$0.033^{+0.033}_{-0.033}$ &
			$0.051^{+0.034}_{-0.034}$ &
			$0.018^{+0.027}_{-0.027}$\\
			$15$--$17$ &
			$-0.023^{+0.025}_{-0.025}$ &
			$0.009^{+0.025}_{-0.025}$ &
			$0.005^{+0.023}_{-0.023}$\\
			$17$--$19$ &
			$-0.030^{+0.033}_{-0.033}$ &
			$0.020^{+0.033}_{-0.033}$ &
			$0.025^{+0.028}_{-0.028}$
			\\
			\bottomrule
		\end{tabular}
	\end{table}
	
	\begin{table}[!tbp]
		\caption{Upper-triangular elements of the experimental covariance matrices for $(A_7,A_8,A_9)$ in each dilepton-mass bin, in units of $10^{-4}$. Cross-bin correlations vanish at the published precision \cite{angular2026data}.}
		\label{tab:cp-cov}
		\centering\small
		\setlength{\tabcolsep}{3pt}
		\begin{tabular}{lrrrrrr}
			\toprule
			$q^2$ bin
			& $V_{77}$ & $V_{78}$ & $V_{79}$
			& $V_{88}$ & $V_{89}$ & $V_{99}$\\
			\midrule
			$1.1$--$2.5$ & 17.640 & -4.685 & 2.673
			& 23.523 & 0.000 & 11.223\\
			$2.5$--$4$ & 20.703 & -2.184 & 1.877
			& 23.040 & -1.440 & 14.063\\
			$4$--$6$ & 10.890 & 0.449 & -1.158
			& 11.560 & -0.459 & 7.290\\
			$15$--$17$ & 6.250 & 1.938 & -0.805
			& 6.250 & -0.345 & 5.290\\
			$17$--$19$ & 10.890 & 2.505 & -0.647
			& 10.890 & -1.201 & 7.840
			\\
			\bottomrule
		\end{tabular}
	\end{table}
	
	\section{Supplementary numerical results and sensitivity tests}
	\label{app:statistics}
	
	For each mediator, the calibration uses 1000 pseudoexperiments at the profiled SM point and 1000 at the best-fit point. The 4000 calibration fits are supplemented by 4000 independent validation pseudoexperiments at eight boundary points. For the temporal phase $\theta_s$, constrained profiling includes both complex solutions of
	\begin{equation}
		(1-\kappa_X w)r_s^2+2r_s+1-w=0,\qquad w=e^{u+i\theta_s},
		\label{eq:quadratic}
	\end{equation}
	with real $u$ profiled continuously. The upper tolerance estimate for the 95th percentile in each generating-point calibration is selected at the binomial order statistic
	$k=1+F^{-1}_{\mathrm{Bin}(1000,0.95)}(1-0.01/16)$,
	accounting for sixteen statistic--generating-point combinations. This ensures 99\% simultaneous Monte Carlo confidence for the upper percentile estimates at the tested points, and is distinct from a simultaneous confidence region in the physics parameters.
	
	\begin{table}[!tbp]
  \caption{Minimum $\chi^2$ for the three data selections with the
    CKM-conjugated amplitudes and the correlated theory uncertainties.}
  \label{tab:fits}
  \centering\small
  \begin{tabular}{lrrr}
  \toprule
  Dataset & $Z'$ & $S_3$ & SM \\
  \midrule
  $\mathcal D_{13}$ & 6.554 & 7.249 & 14.673 \\
  $\mathcal D_{28}$ & 21.918 & 22.524 & 30.016 \\
  $\mathcal D_{68}$ & 73.323 & 70.332 & 80.791 \\
  \bottomrule
  \end{tabular}
\end{table}
	
\begin{figure*}[!t]
		\centering
		\includegraphics[width=.98\textwidth]{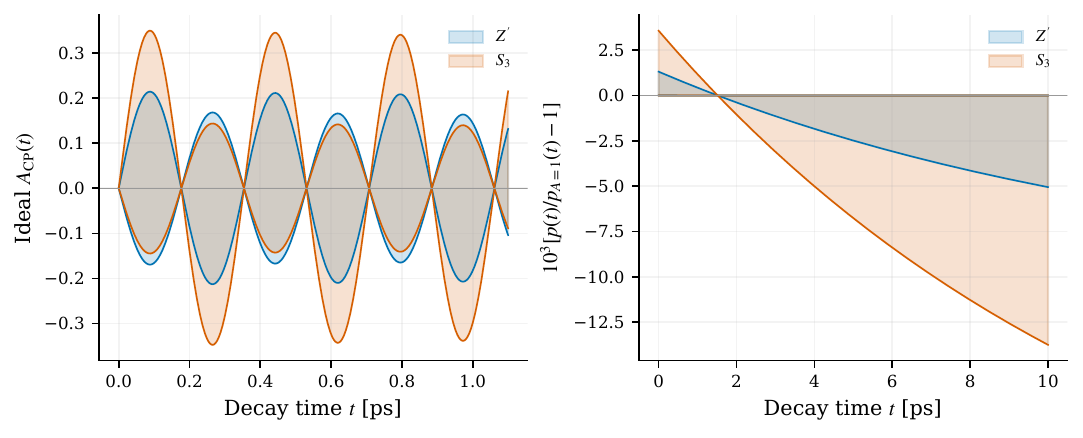}
		\caption{Time-dependent predictions obtained from the calibrated phase regions. Left: ideal tagged CP asymmetry from Eq.~\eqref{eq:tagged}. Right: normalized untagged decay-time distribution relative to $\ADG=1$, using Eq.~\eqref{eq:untagged}. The central values $\Delta m_s=17.766\ \mathrm{ps}^{-1}$, $\Gamma_s=0.6600\ \mathrm{ps}^{-1}$, and $\Delta\Gamma_s=0.0783\ \mathrm{ps}^{-1}$ are used. The bands describe the phase dependence without detector effects.}
		\label{fig:time}
	\end{figure*}

	The time-dependent tagged and untagged distributions corresponding to the calibrated phase regions are shown in Fig.~\ref{fig:time}. The comparison uses central decay widths and does not include detector acceptance or tagging dilution.
	
	\begin{table*}[!tp]
  \caption{Best-fit amplitudes, physical coupling moduli and temporal
    observables from $\mathcal D_{68}$. The quantities $\mathcal C_q$
    are $|g_{qb}|$ for $Z'$ and $|\xi_q|$ for $S_3$.}
  \label{tab:best}
  \centering\small
  \begin{tabular}{lcccccc}
  \toprule
  Model & $r_d$ & $r_s$ & $\mathcal C_d$ & $\mathcal C_s$ & $\Smm^s$ & $\ADG$ \\
  \midrule
  $Z'$ & $-0.0795+0.1052i$ & $-0.0387+0.0173i$ & 0.00475 & 0.00727 & 0.0438 & 0.9990 \\
  $S_3$ & $-0.7849+0.9042i$ & $-0.0365+0.0498i$ & 0.04311 & 0.01065 & 0.1032 & 0.9947 \\
  \bottomrule
  \end{tabular}
\end{table*}
	
	\begin{table}[!tbp]
  \caption{Profile thresholds from the bootstrap with the same
    amplitudes as the data fit. The first three columns refer to
    individual projections, and the last to the complex $r_s$ plane.}
  \label{tab:cutoffs}
  \centering\small
  \setlength{\tabcolsep}{3pt}
  \begin{tabular}{lrrrr}
  \toprule
  Model & $\theta_s$ & $\mathcal C_d$ & $\mathcal C_s$
        & $(\operatorname{Re}r_s,\operatorname{Im}r_s)$ \\
  \midrule
  $Z'$ & 6.317 & 7.469 & 7.959 & 7.959 \\
  $S_3$ & 5.454 & 8.001 & 7.058 & 7.083 \\
  \bottomrule
  \end{tabular}
\end{table}
	
	\begin{table}[!tbp]
  \caption{Acceptance percentages from 500 independent
    pseudoexperiments at each listed boundary point.}
  \label{tab:validation}
  \centering\small
  \setlength{\tabcolsep}{3pt}
  \begin{tabular}{lrrrr}
  \toprule
  Generating point & $\theta_s$ & $\mathcal C_d$ &
    $\mathcal C_s$ & $r_s$ (2D) \\
  \midrule
  $Z'$ $\theta_s^-$ & 95.6 & 99.2 & 99.8 & 95.6 \\
  $Z'$ $\theta_s^+$ & 96.4 & 98.8 & 99.8 & 97.2 \\
  $Z'$ $\mathcal C_d^+$ & 98.4 & 99.4 & 99.6 & 98.2 \\
  $Z'$ $\mathcal C_s^+$ & 99.4 & 99.6 & 99.6 & 99.0 \\
  $S_3$ $\theta_s^-$ & 98.2 & 99.6 & 99.2 & 97.0 \\
  $S_3$ $\theta_s^+$ & 97.4 & 99.4 & 99.0 & 97.2 \\
  $S_3$ $\mathcal C_d^+$ & 99.0 & 99.4 & 99.6 & 98.4 \\
  $S_3$ $\mathcal C_s^+$ & 97.4 & 99.8 & 99.2 & 96.0 \\
  \bottomrule
  \end{tabular}
\end{table}
	
	\begin{table*}[!tp]
		\caption{Sensitivity of the temporal projections to selected variations of the theoretical and experimental inputs. The intervals are obtained from nominal one-dimensional profiles and should not be interpreted as recalibrated confidence regions. These auxiliary variations use the archived angular response and serve as sensitivity checks around its nominal reference fit.}
		\label{tab:stress}
		\centering\small
		\begin{tabular}{lcccc}
			\toprule
			Variation &
			$\Smm^s$ ($Z'$) &
			$\Smm^s$ ($S_3$) &
			min $\ADG$ ($Z'$) &
			min $\ADG$ ($S_3$)\\
			\midrule
			Reference &
			$[-0.122,\,0.178]$ &
			$[-0.106,\,0.311]$ & 0.984 & 0.950\\
			Angular theory errors $\times2$ &
			$[-0.118,\,0.174]$ &
			$[-0.108,\,0.324]$ & 0.985 & 0.946\\
			Angular asymmetric errors &
			$[-0.122,\,0.177]$ &
			$[-0.110,\,0.307]$ & 0.984 & 0.952\\
			Angular correlations removed &
			$[-0.153,\,0.167]$ &
			$[-0.142,\,0.304]$ & 0.986 & 0.953\\
			$R_K$ high-bin correlation $+0.3$ &
			$[-0.102,\,0.171]$ &
			$[-0.106,\,0.311]$ & 0.985 & 0.950\\
			NP mixing $\times0.9$ &
			$[-0.128,\,0.185]$ &
			$[-0.106,\,0.311]$ & 0.983 & 0.950\\
			NP mixing $\times1.1$ &
			$[-0.116,\,0.172]$ &
			$[-0.106,\,0.311]$ & 0.985 & 0.950\\
			$y_{b\mu}=1$ &
			$[-0.122,\,0.178]$ &
			$[-0.110,\,0.317]$ & 0.984 & 0.948\\
			71-observable control &
			$[-0.145,\,0.186]$ &
			$[-0.105,\,0.317]$ & 0.983 & 0.948
			\\
			\bottomrule
		\end{tabular}
	\end{table*}

	The improvements over the SM in $\mathcal D_{68}$ are $7.468$
for $Z'$ and $10.459$ for $S_3$. The corresponding exact $95\%$
binomial intervals for the simulated tail fractions are
$[0.135,\,0.181]$ and $[0.030,\,0.055]$. Pooling the equivalent SM-generating
samples gives a goodness-of-fit tail probability of $0.132$.
The lowest lower endpoint of the $95\%$ binomial coverage intervals
at the tested points is $93.4\%$.

For the observable-by-observable comparison, we use the
principal covariance submatrices and refit both sectors after
each data selection. Starting from $\mathcal D_{28}$ and adding
$S_{2c}$, $S_3$, $S_4$, $S_5$, $A_{\rm FB}$, $S_7$, $S_8$ or
$S_9$ individually gives $\Delta\chi^2_{\rm sec}=$
$17.9$, $5.4$, $29.8$, $230.5$, $150.8$, $15.7$, $6.5$, $5.6$, in that order. Keeping the full angular sample
apart from $S_5$ or $A_{\rm FB}$ gives $286.0$ or $298.9$,
respectively. Dropping both gives $59.2$.
Each number includes a fresh profile over the nuisance parameters.

	The nominal one-dimensional temporal projections are varied across the angular uncertainty model, the covariance, new-physics mixing coefficients, and selected flavor parameters. Table~\ref{tab:stress} reports these comparisons using the same $\Delta\chi^2=3.8415$ threshold rather than separately calibrated cutoffs. Doubling the angular theory-error field modestly enlarges the scalar phase range; removing correlations produces a larger shift. Variations of flavor and mixing inputs, alternative ensemble constructions, and a control with potentially overlapping $R_{K^*}$ data do not alter the qualitative conclusions. The latter control assumes an unavailable cross-analysis covariance and is not included in the primary results.

	\section{Numerical and convention checks}
	\label{app:checks}
	
	The semileptonic and kaon predictions use \texttt{flavio}~2.7.1 and \texttt{wilson}~2.5.2~\cite{flavio2018,wilson2018,kaons2015}. Correlated non-CKM uncertainties in the complementary observables are propagated with a common ensemble of 160 draws. For the angular observables, we use 512 common draws of 19 form-factor and 18 nonlocal-amplitude parameters. If $\bm f^{(s)}$ denotes the vector of predictions for draw $s$ and $\bar{\bm f}$ the ensemble mean, the columns of the angular theory-error matrix are $J_{is}=(f_i^{(s)}-\bar f_i)/\sqrt{511}$. Hence $JJ^T$ is the sample covariance of the predicted observables. For a fixed correlated Gaussian block, writing $\bm v=\bm\mu-\bm d$, stationarity with respect to $\bm\eta$ gives $(\mathbb{1}+J^TV_{\rm exp}^{-1}J)\bm\eta=-J^TV_{\rm exp}^{-1}\bm v$. Substitution into the criterion and the matrix inversion lemma give $\bm v^T(V_{\rm exp}+JJ^T)^{-1}\bm v$.
	
	We reconstruct the quadratic angular numerators and rates at $c=0,\pm1,i$ using 40-point Gauss--Legendre integration, and form the normalized observables afterwards. The angular analysis uses the bias-corrected experimental
	central values and the corresponding covariance matrix,
	preserving the published observable ordering and correlations.
	After applying the experimental unit conversion, we verify
	the symmetry and positive definiteness of the selected
	covariance matrix.
	
	For the conjugate decay we conjugate the weak CKM factors and
the new semileptonic coefficients, keeping the strong amplitudes
fixed. In the low-$q^2$ spectator term the QCDF integrals define
two strong kernels, $T_\lambda$ and $U_\lambda$. We combine them
as $\lambda_t^s T_\lambda+\lambda_u^s U_\lambda$, with
$\lambda_u^s=V_{ub}V_{us}^*$, and conjugate both CKM factors for
the corresponding antiparticle. We also conjugate the CKM
prefactor of the residual nonlocal term. The central angular
response is tabulated at the reference CKM point and at signed
one-standard-deviation displacements for the linear interpolation
used while profiling the four CKM inputs. Direct reevaluation at the
sector minima and the calibrated endpoints changes each angular
observable by less than $1.8\times10^{-3}$ of its experimental
standard deviation. In a separate check with real new-physics
coefficients and $\gamma=10^{-3}$ radians, the CP asymmetries stay
below $3.9\times10^{-6}$ even with complex strong amplitudes.

For the recoil-window check, we keep the form-factor draws and
the central amplitudes and change the $q^2$ dependence of the
residual nonlocal terms. In each helicity we write the effective
shift as
\begin{equation}
  \delta W_\lambda(x)=\sigma_\lambda
  \sum_j\sqrt{w_j(2j+1)}\,P_j(x)\,z_{\lambda j},
  \label{eq:branch_hadronic_variant}
\end{equation}
where $P_j$ are Legendre polynomials, $x$ runs linearly from
$-1$ to $1$ across each recoil window, and the real and imaginary
parts of the $z_{\lambda j}$ are independent standard normal
variables. We use weights $(0.55,0.30,0.15)$ for the effective
dipole shift at low $q^2$ and $(0.65,0.35)$ for the effective
vector shift at high $q^2$. In each window, $\sigma_\lambda$
matches the mean variance of one Cartesian component of the
reference shift across that window. We propagate the 512
predictions through the same covariance and profiling procedure.

	To verify that analytic profiling preserves the fitted criterion, we
	compare the explicit Gaussian-field construction with the profile
	identity, obtaining numerical agreement within
	$1.5\times10^{-14}$ for representative correlated blocks. Checks also
	verify the exact CKM construction, physical coupling conversion,
	Eq.~\eqref{eq:quadratic}, the CP circle, the integrated normalization
	of Eq.~\eqref{eq:untagged}, and the effective lifetime. The parameter configurations at the profile endpoints are checked against the corresponding cutoff and the unrestricted minimum.
	
	Because the selected-data analysis combines Gaussian and polynomial
	response approximations with a finite ensemble and discrete contour
	grids, the bootstrap tests the resulting numerical criterion and
	data-generation prescription. Its interpretation is consequently
	confined to the experimental information and perturbative order
	included in that prescription.
	
	\FloatBarrier

\end{document}